\documentclass[11pt,twoside]{article}
\usepackage{asp2004}
\usepackage{psfig}
\usepackage{epsf}
\usepackage{natbib}
\usepackage{graphics}
\usepackage{lscape}
\def\edcomment#1{\iffalse\marginpar{\raggedright\sl#1\/}\else\relax\fi}
\reversemarginpar

\begin{document}
\title{Effects of Rotation on Presupernovae Models}
\author{Georges Meynet, Raphael Hirschi, \& Andr\'e Maeder}
\affil{Geneva Observatory, CH-1290 Sauverny, Switzerland}

\begin{abstract}
We show that 
rotation strongly affects
the nature of the supernova progenitor (blue/red supergiant or Wolf--Rayet star),
and thus the supernova types. In particular our models well reproduce the variations of
the number ratio SNIb/Ic to SNII with metallicity. Rotation also produces envelope
enrichments of the N/C ratio, and increases the size of the CO cores. We show the evolution
of the specific angular momentum up to the preSN stage and make comparison with neutron stars.
We suggest that the rare WO stars, preferentially formed at low metallicity, are the progenitors
of GRB.
\end{abstract}
\thispagestyle{plain}

\section{Some observational tests of the stellar models}

Models taking account of the effects of rotation for massive stars can reproduce the observed surface enrichments
(\citeauthor{Hel00} \citeyear{Hel00}; \citeauthor{MMV} \citeyear{MMV}),
they can explain the great number of red supergiants observed in the SMC 
\citep{MMVII}. In both cases, non--rotating models fail to fit these observed features.
Rotation also favours the formation of WR stars (\citeauthor{Maeder87} \citeyear{Maeder87};
\citeauthor{Fl95} \citeyear{Fl95};
\citeauthor{MMX} \citeyear{MMX}, \citeyear{MMXI}).
Let us recall that for a star to be considered as a WR star, at least two conditions must be fullfilled: first, the star
must be in the blue part of the HR diagram (typically $\log T_{\rm eff} > 4.0$) and secondly, 
the mass fraction of hydrogen at the surface $X_s$ must be inferior to $\sim$0.4. In the case
of non--rotating models, the decrease of the H--abundance at the surface can only result
from the uncovering of the outer layers by stellar winds. In rotating models, it results from the action of both
the stellar winds and the rotational mixing. Rotation thus adds its effects to that of mass loss and
makes easier the entry into the WR phase.
Numerical models show that,
for a given initial mass and metallicity, rotation increases the WR lifetime with respect
to the values obtained by non--rotating models. It also 
lowers the minimum initial mass of single stars going through a WR phase. In that respect, rotation
has qualitatively similar effects as an enhancement of the mass loss rates.

One can easily estimate the theoretical
number ratio of WR to O--type stars in a region of constant star formation.
This ratio is simply given by the ratio of the averaged lifetimes of a WR star to that
of an OV--type star. The averaged lifetimes are weighted means of the lifetimes over 
the initial mass function (IMF).
Assuming a Salpeter IMF slope (${\rm d}N/{\rm d} M\propto M^{-(1+x)}$, $x$ = 1.35), considering the O--type and WR star lifetimes 
given in \citet{MMXI}, we obtain the predicted ratios plotted in Fig.~1 (left panel).
We assume here that the $\upsilon_{\rm ini}$ = 300 km s$^{-1}$
stellar models are well representative of the behaviour of the majority of the OB stars.
The values for the non--rotating models are well below the observed values. 
The ratios predicted by the models with rotation are
in much better agreement with
the observations.

What do such models, which successfully fit many observed properties of massive stars, 
predict for the presupernovae stage and thus for the initial conditions
of the supernovae explosions ? This is the question addressed in the next sections.

\begin{figure}[!t]
\plottwo{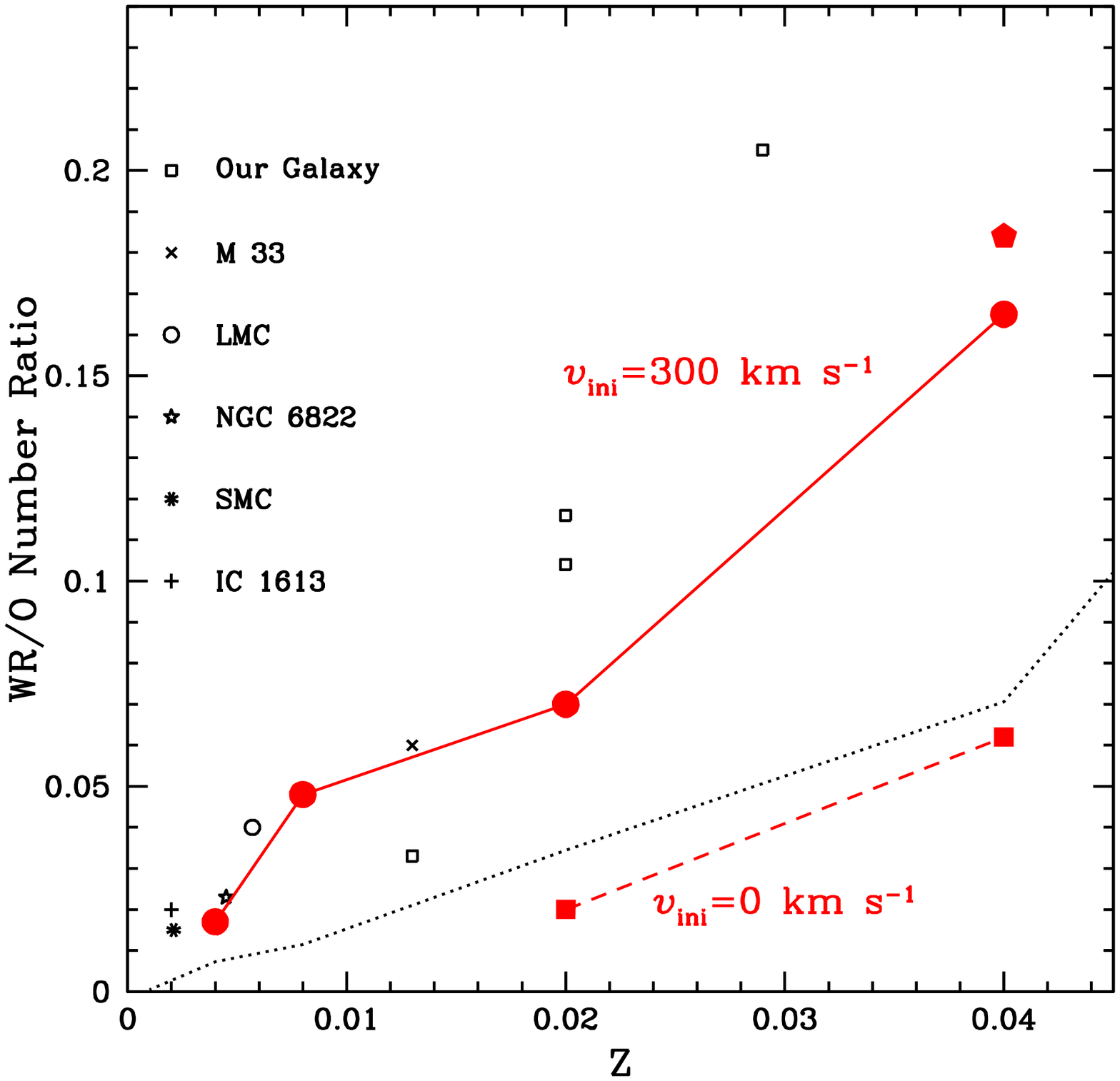}{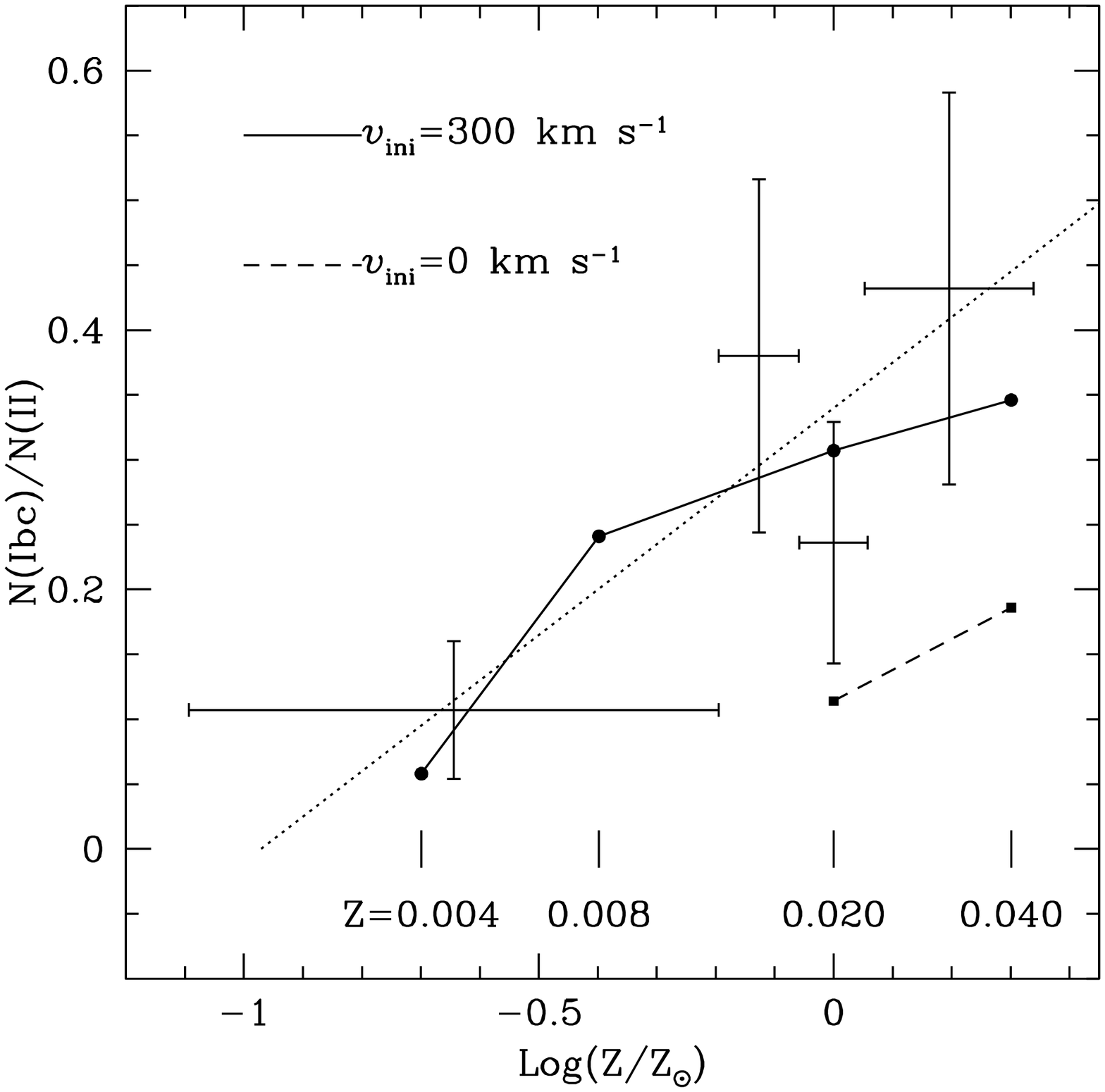}
\caption{{\it Left:} Variation of the number ratios of Wolf--Rayet stars to O--type stars as a function of the metallicity.
The observed points are taken as in  \citet{MM94}.
The dotted line shows the predictions of the models
of \citet{Mey94} with normal  mass loss rates.
The continuous and the dashed lines show the predictions of the present rotating and 
non--rotating stellar models
respectively. The black pentagon shows the ratio predicted by Z=0.040 models computed
with a metallicity dependence of the mass loss rates during the WR phase. {\it Right:} Variation of the number ratios of type Ib/Ic supernovae to type II supernovae. The crosses
  with the error bars correspond to the values deduced from observations by \citet{Pr03}. The
  dotted line is an analytical fit proposed by these authors. The continuous and dashed line show the predictions of 
  the present rotating and non--rotating stellar models.
}
\end{figure}

\section{Nature of the supernova progenitors}

Depending on rotation, a given initial mass star may end its life as a red, a blue supergiant or as a Wolf--Rayet star.
Let us first consider this latter case.
Current wisdom associates the supernovae of type Ib/Ic with the explosion of WR stars, the
H--rich envelope of which has been completely removed either by stellar winds or by mass transfer 
through Roche Lobe overflow in a close binary system. 
For single star models at least, 
theory predicts that the fraction of type Ib/Ic supernovae
with respect to type II supernovae should be higher at higher metallicity. The reason is the same
as the one invoked to explain the increasing number ratio of WR to O--type stars with the metallicity, $Z$, namely the
growth of the mass loss rates with $Z$. \citet{Pr03}
have derived from published data the
observed number ratios of type Ib/Ic supernovae to
type II supernovae for different metallicities. The regions considered  are regions of constant star formation rate. 
Their results are plotted in the right panel of Fig.~1 . One sees that rotating models
give a much better fit to the observed data than non--rotating ones. This comparison can be viewed as a check
of the lower initial mass limit M$_{\rm WNE}$ of the stars evolving into a WR phase without hydrogen.
The comparison between the observed and
predicted number ratio of WR to O--type stars shown in the left panel of Fig.~1 involves not only the value of the minimum 
initial mass of stars evolving into the WR phase but also the durations of the WR phase. 
In that respect the comparison with the supernovae ratios is a more direct check of the correctness of the value of $M_{\rm WNE}$
which at high metallicity appears rather close to M$_{\rm WR}$.

\begin{figure}[!t]
\plottwo{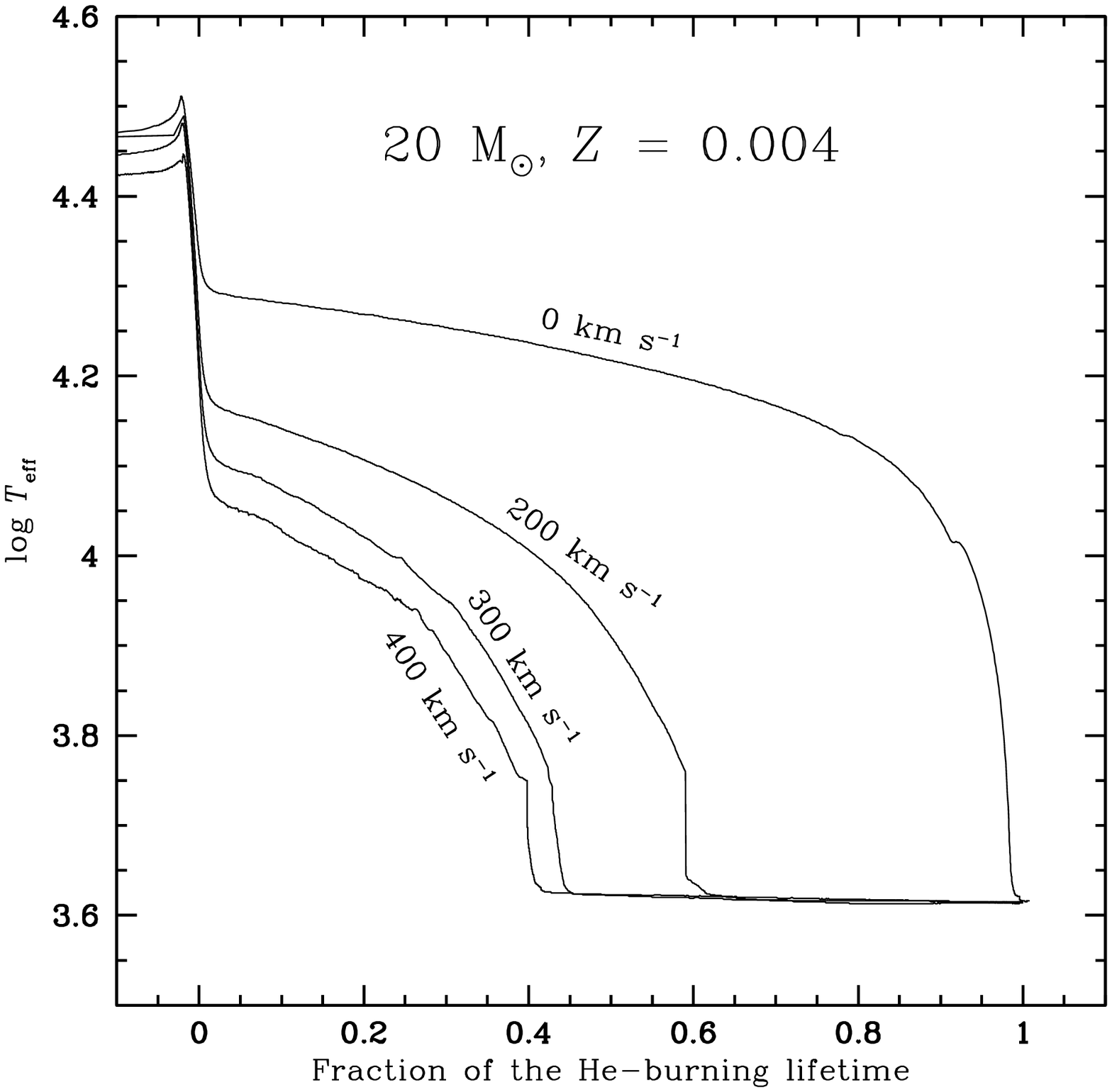}{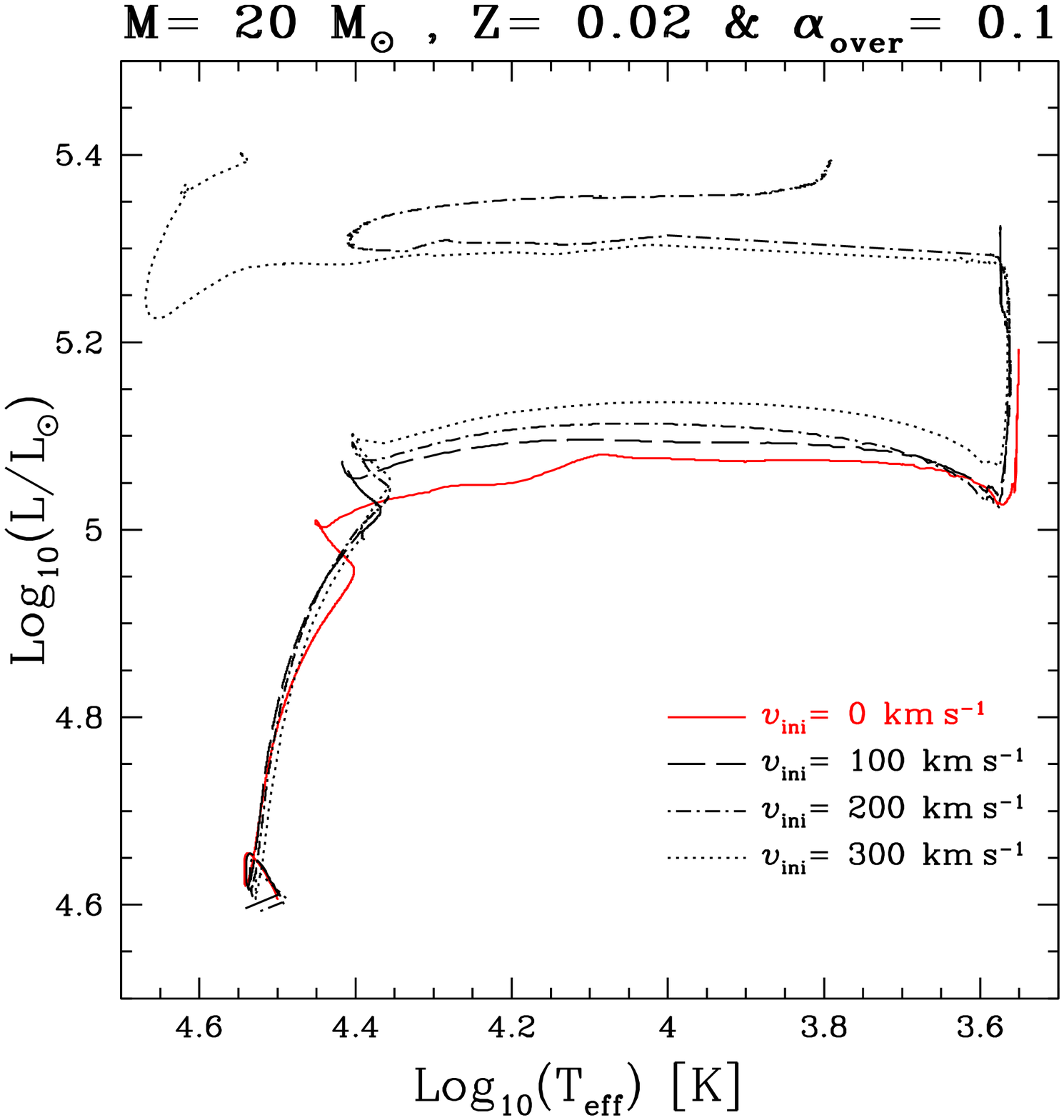}
\caption{{\it Left:} Evolution of the $T_{\mathrm{eff}}$
as a function of the fraction of the lifetime spent
in the He--burning phase for 20 M$_\odot$ stars with different
initial velocities. {\it Right :} HR--diagram 
for 20 $M_{\sun}$ models at solar metallicity: solid, dashed, dotted-dashed and dotted lines correspond 
respectively to
$v_{\rm{ini}}$= 0, 100, 200 and 300 km\,s$^{-1}$.
}
\end{figure}

Rotation also affects the colour of the supergiants.
In Fig.~2 (left), the evolution
of $\log T_{\rm eff}$ as a function of time during the core He--burning phase is shown. One sees that
at $Z=0.004$, non--rotating models spend nearly the whole He--burning phase in the blue part of the HR diagram, while, when
rotation increases, the time spent in the red increases at the expense of the time spent in the blue.
The physical reason for this behaviour is given in \citet{MMVII} and is related to the mixing induced by rotation.
Interestingly, as briefly mentioned in Sect.~1, observation shows a great number of red supergiant in the Small
Magellanic Cloud, a fact which may be easily reproduced by rotating models but which is in contradiction
with the predictions of the non--rotating ones.

Figure~2 (right panel) illustrates this same behaviour 
for different 20 M$_\odot$ models
at solar metallicity. For $\upsilon_{\rm ini} \le ~100$, the stellar models
would explode when the star is a red supergiant. For higher initial velocities, the supernova progenitor
would be a blue supergiant or even a WR star. Not only
the colours of the supergiants depend on rotation but
also their structure and their chemical composition 
(\citeauthor{HLW00} \citeyear{HLW00}; \citeauthor{HMMXII} \citeyear{HMMXII}).
For instance the mass of the CO core in the $\upsilon_{\rm ini}$ = 300 km s$^{-1}$ model
is 5.86 M$_\odot$, which is about 1.5 times the mass of the corresponding core in the non--rotating model.
Also the N/C ratio at the surface of the non--rotating model at the presupernova stage is equal to
3.4 (in number). At the surface of the rotating models the N/C ratios are equal to
8.2, 40.2 and 76.0 for $\upsilon_{\rm ini}$ = 100, 200 and 300 km s$^{-1}$ respectively.
These differences will modify the supernova explosion, the chemical composition of the ejecta and
may change the nature of the stellar remnant at least for those stars which are at the limit
between those producing neutron stars and those giving birth to black holes.

\section{Angular momentum at the pre-supernova stage and the progenitors of GRBs}

\begin{figure}[!t]
\plottwo{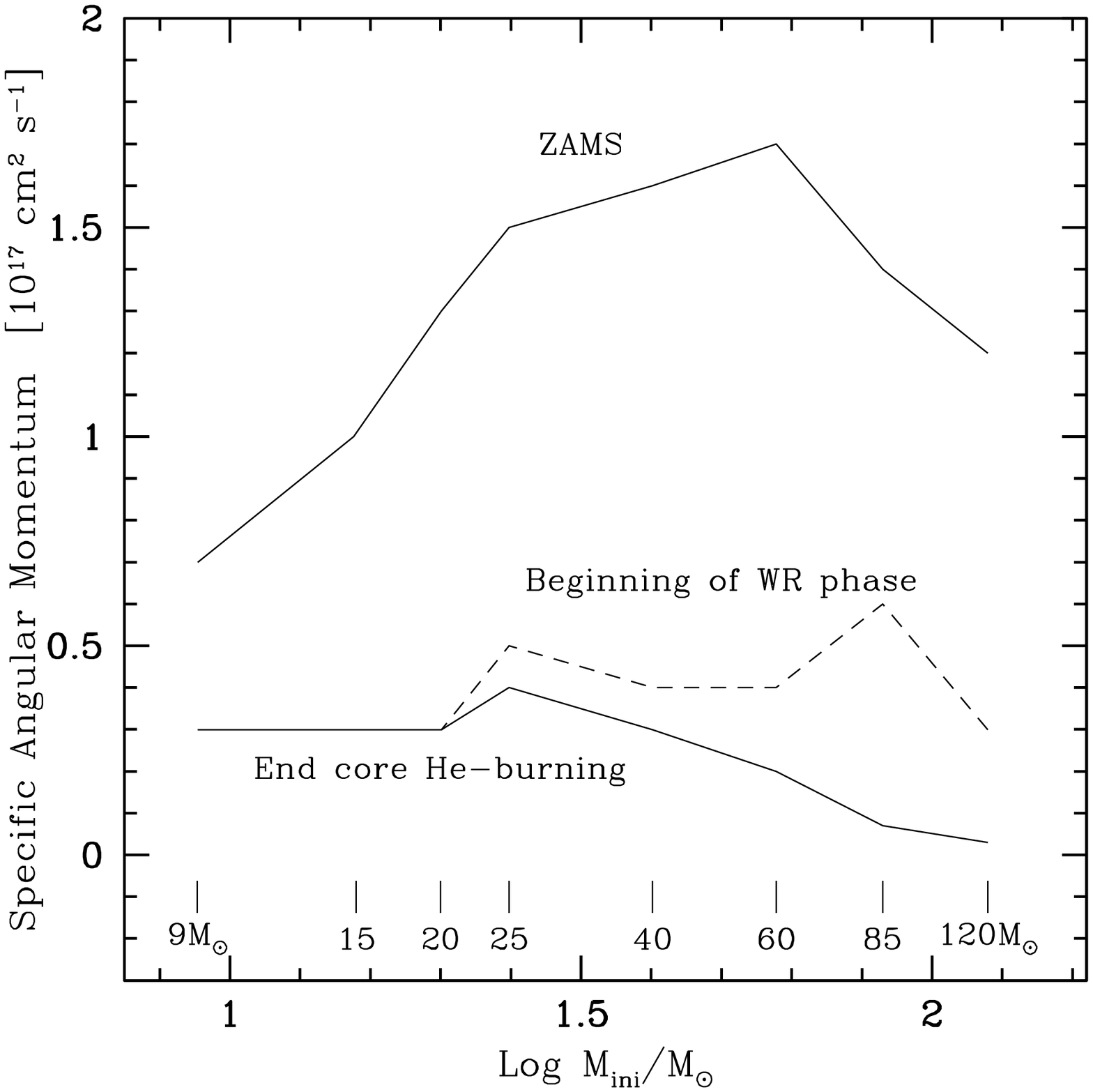}{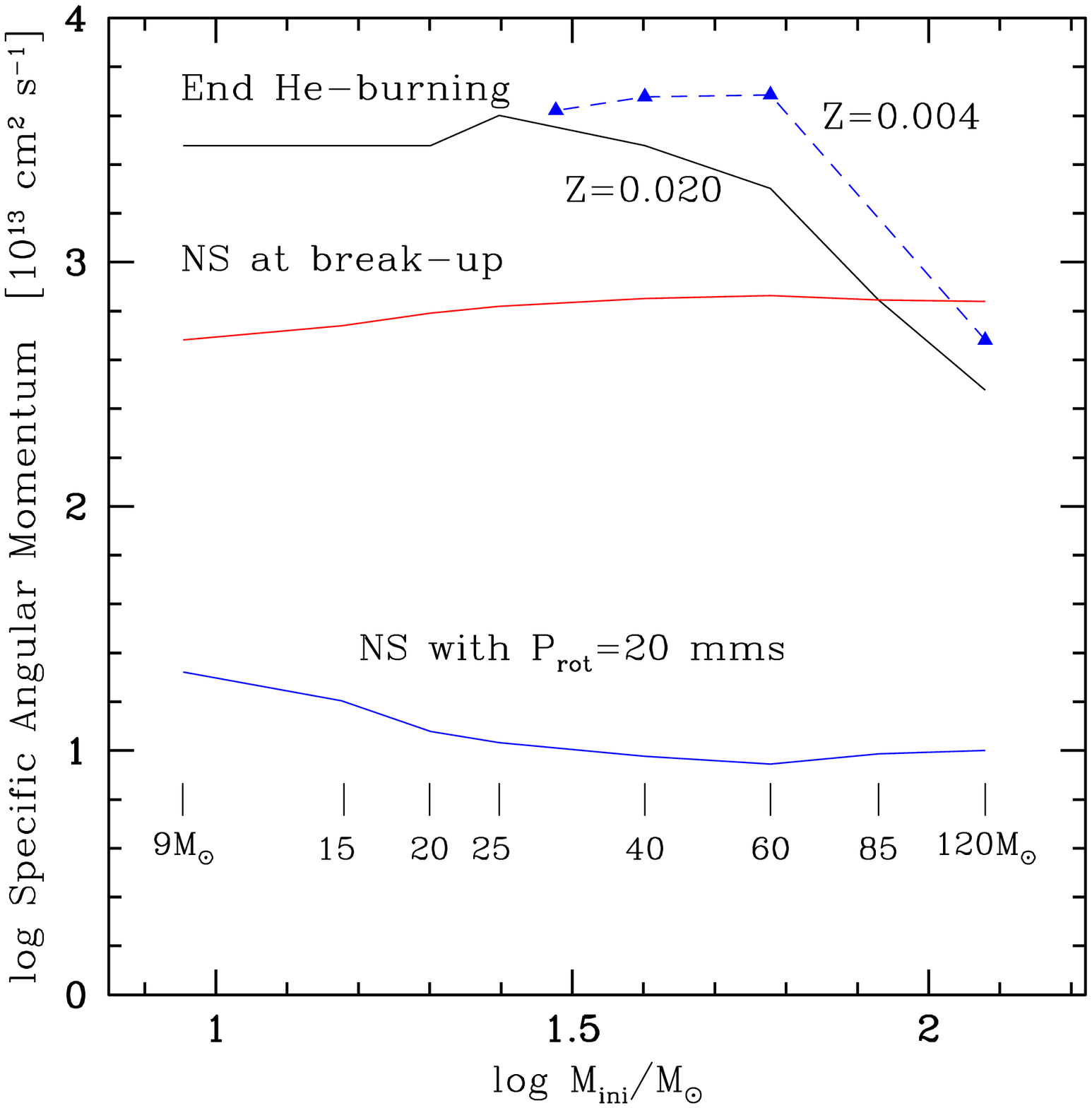}
\caption{{\it Left :} Specific angular momentum in the central regions (see text) on 
the ZAMS, at the entrance into the WR phase, and at the end of the core He--burning phase.
{\it Right :} Specific angular momentum in the central regions at the end of the core He--burning phase,
for neutron stars rotating at the break--up limit and for young pulsars.
For the momentum of inertia of the neutron star we used the relation $I_{\rm NS}=
0.32\  M R^2$ as in \citet{he98}, where $M$ is taken equal to $M_{\rm rem}$
and R is given by the relation $R/R_\odot=15.2/(M/M_\odot)^{1/3}$ derived from
the theory of polytropes.}
\end{figure}

Meridional circulation, shear instabilities, convection, radial mouvements, mass loss at the surface, all
these processes affect the distribution of the angular velocity in the stellar interior and therefore the
distribution of the angular momentum. 
For the most massive stars, the evolution of the total angular momentum is dominated by mass loss. Typically for the 60 M$_\odot$ model with
$\upsilon_{\rm ini}$ = 300 km s$^{-1}$, the total angular momentum decreases by a factor 6 during the O--type star phase
and by a factor 17 during the WR phase. This star looses more than 99\% of its initial angular momentum, 
between the ZAMS and the end of the core He--burning phase.
For the 120 M$_\odot$, the quantity of angular momentum lost amounts
to 99.96\% of its initial value. When the initial mass decreases, mass loss rates become weaker and thus
the quantities of angular momentum lost are smaller.

More interesting is how the angular momentum of the core of the star is evolving as a function of time.
The angular momentum of the core depends on the
transport processes active during the whole evolution of the star. Among the three
transport processes considered, convection, shear diffusion and meridional circulation, 
the first two always transport angular 
momentum inside--out, only meridional circulation can, in some circumstances, transport 
angular momentum from the outer parts of the star into its inner parts. 
In Fig.~3 (left panel) the specific angular momentum of the central region of the star is shown for
various initial masses at solar metallicity and for three evolutionary stages: the ZAMS, the beginning
of the WR phase for the most massive stellar models, and the end of the core He--burning phase.
By central region of the star, we mean here the part of the star which at the end will remain locked
in the stellar remnant (either under the form of a black hole or a neutron star\footnote{The masses of the remnants are deduced from the CO core masses 
as in \citet{Ma92}.}).
One sees that the angular momentum of the core decreases as a function of time, clearly
showing that the outwards transport processes dominate over the inwards ones.
The most important decrease occurs during the core H--burning phase.

The right panel of Fig.~3 compares the specific angular momentum of neutron star to that
at the end of the core He--burning phase. At this stage,
the specific angular momenta of the central regions have nearly reached their final values (see the
contribution by Hirschi in the present volume). We see that
angular momentum contained in the core is much higher than the one contained in young
pulsars. This important excess of angular momentum might be evacuated either during
the pre--supernova evolution by mechanisms not accounted for in the present models
(for instance the effects of a magnetic field have been explored by 
\citeauthor{WH04} \citeyear{WH04} and \citeauthor{MagII} \citeyear{MagII}), or at the time
of the supernova explosion, or at the beginning of the neutron star life.
On the other hand, the specific angular momentum obtained is sufficiently high
for generating a GRB through the collapsar model \citep{Wo93}. 
A high angular momentum is likely not the only condition however for producing a GRB, 
the star should form a black hole and should likely 
have lost its H--rich envelope. 

The association of GRB with hypernovae of the class of SNIc
is supported by several observations 
(e.g. see \citeauthor{Mazzali03} \citeyear{Mazzali03}). SNIc result from the 
explosion of a star without H and with little or no He. 
This corresponds to a rare category of WR stars:
the so--called WO stars. These stars show the products of He-burning, with an excess of C+O with respect to He and
O $>$ C. They result from the 
evolution of stars  with M $\geq$ 60 $M_{\odot}$ 
{\emph{at low metallicity only}} \citep{SmithM91}. Why only at low metallicity ?
The reason is the following: at high $Z$, mass loss is high, thus when the products
of the 3$\alpha$ reaction appear at the surface, they are in an early stage of nuclear processing,
i.e. with a low (C+O)/He ratio. At low $Z$, because the stellar winds are weaker,
the products of the 3$\alpha$ reaction rarely appear at the stellar surface
and if they do it (e.g. in case of high rotation) this  occurs very late in the evolution,
i.e. when (C+O)/He  is high and the star is an early WC star or a WO star. Such a scenario is confirmed by the observation
which shows that early WC 
types and WO stars are only found in lower $Z$ regions \citep{SmithM91}.
At solar metallicity, the present stellar models predict no WO stars to be formed.
At $Z$ = 0.004, only stars in the mass range between $\sim$50--70 M$_\odot$ become WO stars.
This would mean that at this metallicity, about 2\% of the core collapse supernovae would
produce a GRB, a value not too far from the estimate deduced from the observed frequency
of GRB \citep{WH04}.

\section{Conclusion}

The inclusion of rotation in stellar models improve the agreement with the observations.
The stellar type of the progenitors of the supernovae, the chemical composition of the ejecta and the
nature of the remnant depend on rotation. The quantity of angular momentum in the central region
at the time of the explosion has an impact on the observed characteristics of the explosion.
Even for stars losing mass at a high rate, the specific angular momentum of the core
is largely sufficient  for producing a collapsar. A GRB, however will only appear when special conditions
are met, realised 
only in a small fraction of core collapse supernovae.
Let us add also that rotation modifies the way massive stars lose mass. It induces anisotropy 
\citep{Ma99} and
enhances the mass loss rates \citep{MMVI}. This may change the conditions in the surroundings of the star
and therefore the interactions of the supernova ejecta with the circumstellar material.


\begin{thebibliography}{}


\bibitem[Fliegner \& Langer(1995)]{Fl95}
Fliegner, J., Langer, N. 1995. 
in IAU Symp. 163, K. A van der Hucht \& P.M. Williams (eds.), (Dordrecht: Kluwer), 326

\bibitem[Heger(1998)]{he98}
Heger, A. 1998, phD, Max--Planck--Institut f\"ur Astrophysik, M\"unchen

\bibitem[Heger \& Langer(2000)]{Hel00}
Heger, A., \& Langer, N. 2000, \apj, 544, 1016

\bibitem[Heger et al.(2000)Heger, Langer, \& Woosley]{HLW00}
Heger, A., Langer, N., Woosley, S.E. 2000, \apj, 528, 368

\bibitem[Hirschi et al.(2004)]{HMMXII}
Hirschi, R., Meynet, G., \& Maeder, A. 2004, A\&A, in press

\bibitem[Maeder(1987)]{Maeder87}
Maeder, A. 1987, \aap, 178, 159

\bibitem[Maeder(1992)]{Ma92}
Maeder, A. 1992, \aap, 264, 105

\bibitem[Maeder(1999)]{Ma99}
Maeder, A. 1999, \aap, 347, 185

\bibitem[Maeder \& Meynet(1994)]{MM94}
Maeder, A., \& Meynet, G. 1994, \aap, 287, 803

\bibitem[Maeder \& Meynet(2000)]{MMVI}
Maeder, A., \& Meynet, G. 2000, \aap, 361, 159, (Paper VI)

\bibitem[Maeder \& Meynet(2001)]{MMVII}
Maeder, A., \& Meynet, G. 2001, \aap, 373, 555, (Paper VII)

\bibitem[Maeder \& Meynet(2004)]{MagII}
Maeder, A., \& Meynet, G. 2004, \aap, 422

\bibitem[Mazzali et al.(2003)]{Mazzali03}
Mazzali, P.A., Iwamoto, K., \& Nomoto, K. 2003, ApJ 599, L95

\bibitem[Meynet \& Maeder(2000)]{MMV}
Meynet, G., \& Maeder, A. 2000, \aap 361, 101, (Paper V)

\bibitem[Meynet \& Maeder(2003)]{MMX}
Meynet, G., \& Maeder, A. 2003, \aap 404, 975, (Paper X)

\bibitem[Meynet \& Maeder(2004)]{MMXI}
Meynet, G., \& Maeder, A. 2004, \aap, in press, (Paper XI)

\bibitem[Meynet et al(1994)]{Mey94}
Meynet, G., Maeder, A., Schaller,
G., Schaerer, D., \& Charbonnel, C. 1994, A\&AS, 103, 97


\bibitem[Prantzos \& Boissier(2003)]{Pr03}
Prantzos, N., Boissier, S. 2003, \aap, 406, 259

\bibitem[Smith \& Maeder(1991)]{SmithM91}
Smith, L.F., Maeder, A. 1991, A\&A, 241, 77

\bibitem[Woosley(1993)]{Wo93}
Woosley, S.E. 1993, ApJ, 405, 273

\bibitem[Woosley \& Heger(2004)]{WH04}
Woosley, S.E., Heger, A. 2004, in IAU Symp.215, A. Maeder \& P. Eenens (Eds.),
in press.
\end{thebibliography}
\end{document}